\documentclass[twocolumn,aps,prl,superscriptaddress,citeautoscript,nofootinbib,showpacs,floatfix]{revtex4}
\usepackage{graphicx}
\usepackage{amsmath}
\newcommand{\BaFeCoAs}{Ba(Fe$_{1-x}$Co$_x$)$_2$As$_2$}
\begin{document}

\title{Local measurement of the superfluid density in the pnictide superconductor Ba(Fe$_{1-x}$Co$_{x}$)$_2$As$_2$ across the superconducting dome} 

\author{Lan Luan}
\author{Thomas M. Lippman}
\author{Clifford W. Hicks}
\altaffiliation{Current address: School of Physics and Astronomy, University of St Andrews, North Haugh, St Andrews, Fife KY16 9SS}
\author{Julie A. Bert}
\author{Ophir M. Auslaender}
\altaffiliation{Current address: Physics Department, Technion-Israel Institute of Technology, Haifa 32000, Israel}
\author{Jiun-Haw Chu}
\author{James G. Analytis}
\author{Ian R. Fisher}
\author{Kathryn A. Moler}
\email[Corresponding author: ]{kmoler@stanford.edu}
\affiliation{Stanford Institute for Materials and Energy Science, SLAC National Accelerator Laboratory, 2575 Sand Hill Road, Menlo Park, CA 94025 and Geballe Laboratory for Advanced Materials, Stanford University, Stanford, CA 94305} 

\begin{abstract}
We measure the penetration depth $\lambda_{ab}(T)$ in Ba(Fe$_{1-x}$Co$_{x}$)$_2$As$_2$ using local techniques that do not average over the sample. The superfluid density $\rho_s(T)\equiv1/\lambda_{ab}(T)^2$ has three main features. First, $\rho_s(T=0)$ falls sharply on the underdoped side of the dome. Second, $\lambda_{ab}(T)$ is flat at low $T$ at optimal doping, indicating fully gapped superconductivity, but varies more strongly in underdoped and overdoped samples, consistent with either a power law or a small second gap. Third, $\rho_s(T)$ varies steeply near $T_c$ for optimal and underdoping. These observations are consistent with an interplay between magnetic and superconducting phases.
\end{abstract}


\maketitle

The magnetic penetration depth $\lambda$ is one of the most important length scales in superconductors \cite{tinkham_introduction_1975}. Its temperature evolution is a sensitive probe of the superconducting gap structure \cite{hardy_precision_1993, prozorov_magnetic_2006} and its value is related to the density of electrons in the superconducting state, $\rho_s(T) \equiv 1/\lambda(T)^2$. Comparing $\rho_s(T)$ from samples across the phase diagram of a family of iron pnictide superconductors can shed light on the pairing mechanism \cite{mazin_unconventional_2008} as well as on the relation between superconductivity and adjacent phases \cite{Vorontsov_interplay_2009}.  The family of \BaFeCoAs\ is a good candidate for such studies because single crystals can be grown cleanly with relatively sharp superconducting transitions \cite{ni_effects_2008, chu_determination_2009}, because the magnetic and structural transitions are well characterized \cite{Nandi_Suppression_2010, Lester_Neutron_2009, Pratt_Coexistence_2009}, and because other careful studies of quasi-particle excitation measurements such as thermal conductivity \cite{Tanatar_HeatTransport_2010, Reid_caxisheat_2010} and Raman scattering \cite{Muschler_Band_2009} have been successfully performed across the phase diagram.

$\lambda$ is difficult to measure accurately. In the pnictides, the proximity of the magnetic phase to superconductivity on the underdoped (UD) side of the superconducting dome prevents measurement of $\lambda$ by $\mu$SR \cite{williams_Superfluid_2009}. Bulk measurements by microwave- and RF-based techniques are made difficult by complex sample topography and inhomogeneity, which can explain the significantly different results among nominally similar samples of both $\Delta\lambda\equiv\lambda(T)-\lambda(0)$ \cite{hashimoto_microwave_2009, hashimoto_Linenodes_2010, Malone_SmFeASO_2009, martin_nonexponential_2009, gordon_london_2009} and $\lambda$ \cite{gordon_doping_2010} measurements. Local-probe studies of $\rho_s(T)$ across the doping range are strongly desirable, because such measurements can obtain $\lambda$ even when the magnetic order is adjacent or co-existing \cite{Luan_local_2010}, and can reduce the error from topography and inhomogeneity \cite{hicks_evidence_2009}. 

In this Letter, we report local measurements of $\lambda_{ab}(T)$, the penetration depth for screening currents flowing in the $ab$ plane in a set of Ba(Fe$_{1-x}$Co$_{x}$)$_2$As$_2$ single crystals grown from self-flux \cite{chu_determination_2009}. We measure at Co composition $x$ across the superconducting dome: underdoped (UD) $x=0.045$, $0.049$, $0.051$; optimally doped (OptD) $x=0.07$; and overdoped (OD) $x=0.085$, $0.11$ \footnote{For UD samples, $\lambda_a\neq\lambda_b$, our measurements give $\lambda_{ab}\equiv\sqrt{\lambda_a\lambda_b}$}. Our measurements average over a few microns. The positions we choose to measure show strong, uniform diamagnetic response, and are at least 15 microns away from topographic steps larger than 0.5 micron. We are able to resolve well-formed vortices to rule out granularity on sub-micron scales except at $x=4.5\%$. We observe systematic evolution of $\rho_s(T)$ with $x$. Our observations suggest strong correlation between magnetism and superconductivity.

We use magnetic force microscopy (MFM) to measure $\lambda_{ab}(T)$ and $\Delta\lambda_{ab}(T)\equiv\lambda_{ab}(T)-\lambda_{ab}(0)$ irrespective of adjacent magnetic states from 5~K to the superconducting transition temperature $T_c$ \cite{Luan_local_2010}. We also use scanning Superconducting QUantum Interference Device susceptometry (SSS) in a $^3$He cryostat to measure $\Delta\lambda_{ab}(T)$ from 0.4~K to 7~K \cite{hicks_evidence_2009}. Both techniques measure the diamagnetic response to a local field source (either the MFM tip or the SQUID field coil) in the Meissner state. This is measured by SSS through the mutual inductance between the field coil and the pickup loop, and by MFM through the derivative of the vertical component of the force between the magnetic tip and the sample $\partial F_z/\partial z$, where $z$ is the tip-sample distance. The diamagnetic response can be approximated as a time-reversed mirror of the source reflected about a plane $\lambda_{ab}$ below the surface when the field source is much further than $\lambda_{ab}$ above the surface \cite{Kogan_Meissner_2003}. Under this approximation, changes in $z$ and $\lambda$ are equivalent, allowing model independent measurement of $\Delta\lambda_{ab}(T)$. By properly modeling the MFM tip-superconductor interaction, we also obtain $\lambda_{ab}(T)$. More details on the techniques are provided in previous publications \cite{hicks_evidence_2009, Luan_local_2010}. 

\begin{figure}
\includegraphics[width=3.1in]{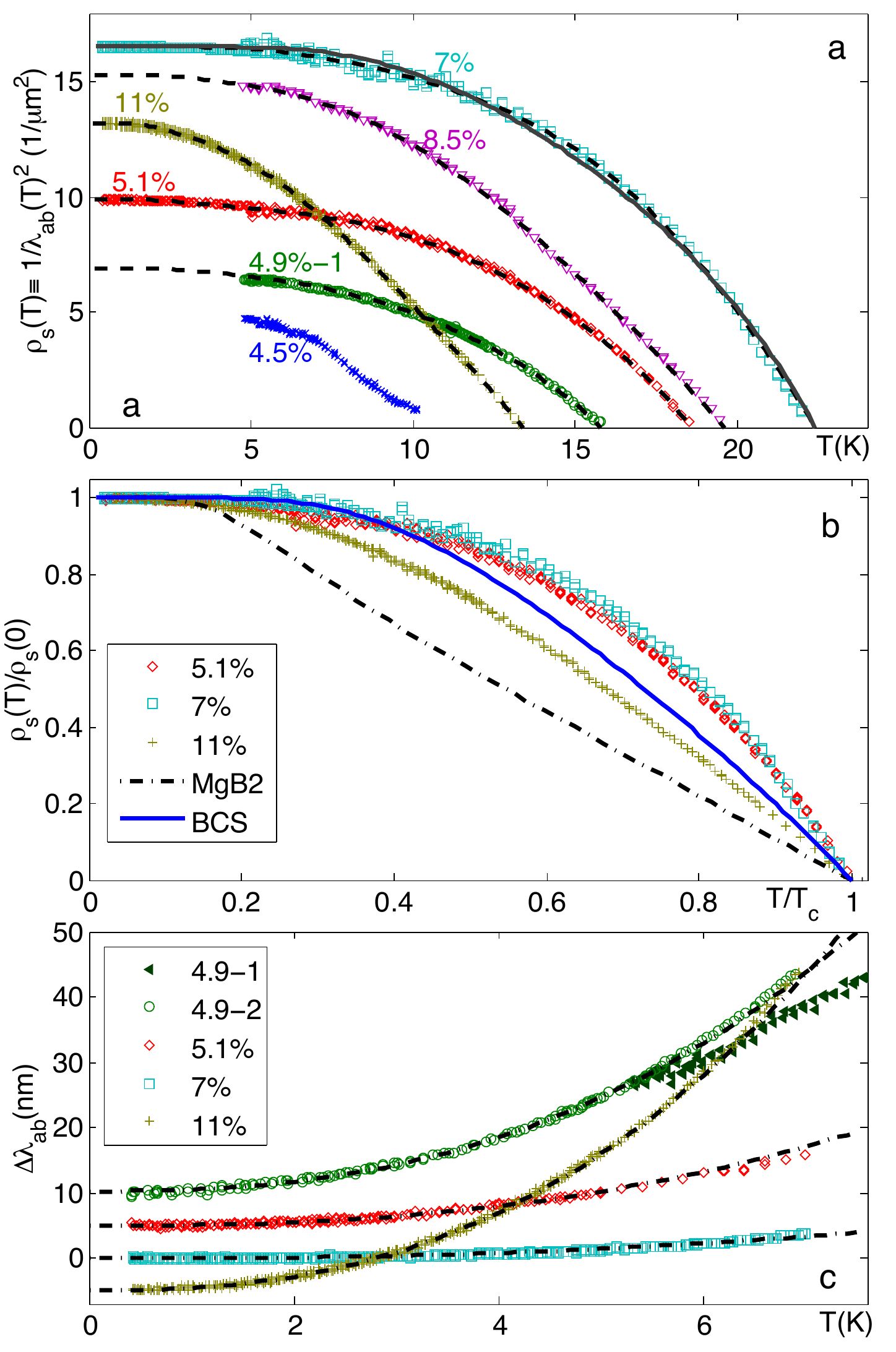}
\caption{\label{fig_penetration} The temperature dependence of $\rho_s(T)\equiv1/\lambda_{ab}(T)^2$ and $\Delta\lambda_{ab}(T)$ shows three systematic trends. %
{\bf a:} $\rho_s(T)$ vs $T$. Solid line: a single-gap equation (Eq.~1) fits only $x=7\%$. Dashed lines: phenomenological two-gap fits (Eq.~2) work well at every doping.  Fit parameters in Table.~\ref{table_para}). At $x=8.5\%$, we did not measure $\lambda_{ab}(T)$, so we offset $\Delta\lambda_{ab}(T)$ by the mean of $\lambda_{ab}(0)$ at $x=7\%$ and $x=11\%$. %
{\bf b:} $\rho_s(T)/\rho_s(0)$ vs $T/T_c$. At $x=5.1\%$(UD) and $x=7\%$(OptD), $\rho_s(T)$ rises more sharply than for MgB$_2$ from Ref.~\cite{Fletcher_MgB2_2005}, single-band weakly coupled BCS theory, or $x=11\%$(OD). %
{\bf c:} $\Delta\lambda_{ab}(T)$ down to 0.4~K measured by SSS at the indicated $x$. Dashed lines: power-law fits (Eq.~3) describe the data well up to 0.3$T_c$ with the power $n$ fixed at 2.5 (fit parameters in Table.~\ref{table_para}). The amplitude $A$ increases away from optimal doping, where $\Delta\lambda_{ab}(T)$ is so flat as to be consistent with exponential behavior. Successive data sets are offset vertically by 5 nm for clarity. $\Delta\lambda_{ab}(T)$ at $x=4.9\%$ measured on a different sample by MFM at $T>5$~K is also plotted. %
}
\end{figure}

We determine $\lambda_{ab}(T)$ from $5$~K to $T_c$ by MFM within $15\%$ error, mostly from uncertainty in the MFM tip geometry. We use the same MFM tip at $x=4.5\%$, $7\%$ and $11\%$, reducing the relative error of $\lambda_{ab}(T)$ among these samples. $\Delta\lambda_{ab}(T)$ by both techniques has $7\%$ error, mostly from calibration uncertainty of the scanner. At most $x$, we measure at least two samples from the same growth batch, one by MFM and one by SSS. When $\Delta\lambda_{ab}(T)$ measured by both the techniques overlaps in the common temperature range, we offset $\Delta\lambda_{ab}(T)$ from SSS by $\lambda_{ab}(T=5K)$ from MFM, and obtain $\rho_s(T)\equiv1/\lambda_{ab}(T)^2$ over the full temperature range, as shown in Fig.~\ref{fig_penetration}. However, $\Delta\lambda_{ab}(T)$ measured by the two techniques do not agree at $x=4.9\%$, presumably due to sample variations. At $x=11\%$, we observe different $\Delta\lambda_{ab}(T)$ by SSS at three locations separated by hundreds of microns (shown in Fig.~\ref{fig_uniformity}), one of which matches the MFM measurement. 

\begin{figure}[t]
\includegraphics[width=3.1in]{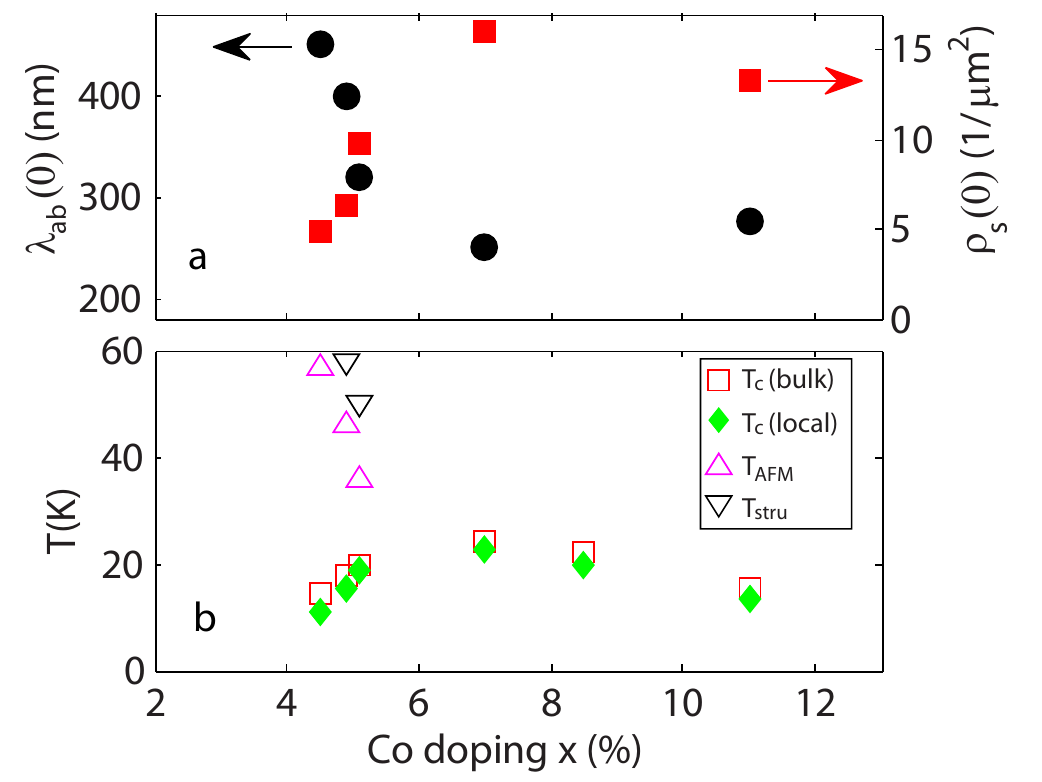}
\caption{\label{fig_para} 
{\bf a:} $\lambda_{ab}(0)$ and $\rho_s(0)$ as extrapolated from Fig~\ref{fig_penetration}(a), showing the rapid drop of $\rho_s(0)$ on the underdoped side. For $x=4.5\%$,  $\rho_s(5K)$ is plotted because the extrapolation is unclear. %
{\bf b:}  Temperatures of structural, magnetic and superconducting transitions $T_{stru}$, $T_{AFM}$ and $T_c$(bulk), from resistivity measurements of samples from the same growth as studied here. $T_c$(local) is from linear extrapolation of $\rho_s(T)$ to zero from the MFM data presented here.}
\end{figure}

\begin{figure}
\includegraphics[width=3.3in]{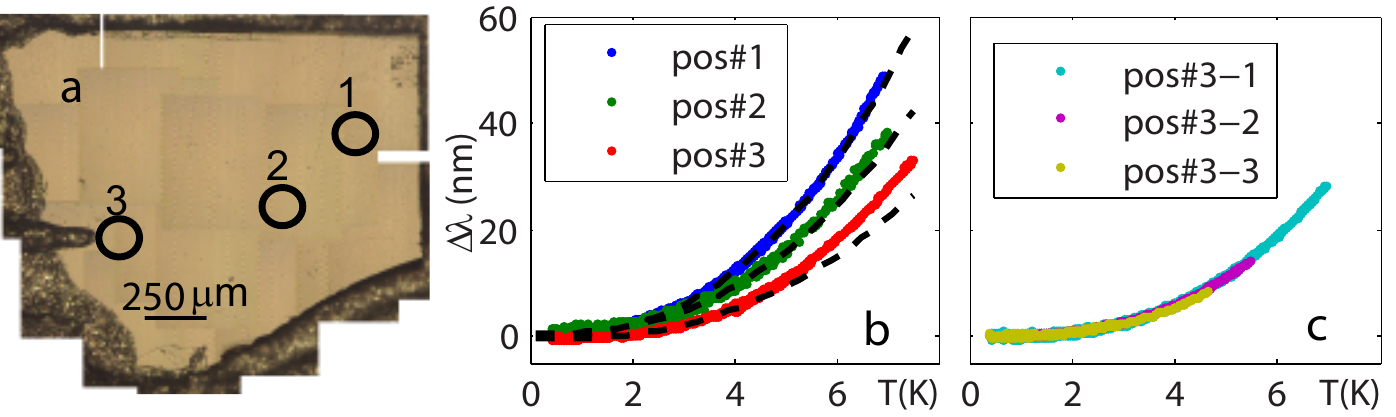}
\caption{\label{fig_uniformity}Inhomogeneity in $\Delta\lambda_{ab}(T)$ observed in an $x=11\%$ sample.
{\bf a:} Photograph of the $x=11\%$ sample measured by SSS. The label of positions corresponds to the position where $\Delta\lambda_{ab}(T)$ was measured. %
{\bf b,c:} $\Delta\lambda_{ab}(T)$ by SSS at positions 1-3 (b) and three different positions around position 3 each separated by $15\mu$m (c). Dashed lines in (b) plot power-law fits with parameters given in Table~\ref{table_para}.}
\end{figure}

We observe a systematic change of $\rho_s(T)$ with Co doping $x$ across the superconducting dome that can be characterized by three trends. First, the zero temperature value $\rho_s(0)$ (Fig.~\ref{fig_para}) peaks at OptD. It is strongly reduced in the UD regime, falling sharply as the magnetic order onsets.  For UD samples, $\rho_s(0)$ falls more quickly than $T_c$, while on the OD side, $\rho_s(0)$ falls less rapidly than $T_c$. This observation is different from a previous measurement by bulk technique that reported $\rho_s(0)$ increasing with doping across the doping dome \cite{gordon_doping_2010}. 

Second, the magnitude of $\Delta\lambda_{ab}(T)$ [Fig.~\ref{fig_penetration}c] at low $T$ increases away from OptD on both sides of the dome. At $x=7\%$, $\Delta\lambda_{ab}(T)$ remains flat within 0.5 nm below 3 K, indicating fully gapped superconductivity,  consistent with the proposed order parameter $s\pm$ \cite{mazin_unconventional_2008}.  We use a phenomenological single-gap fit to $\rho_s(T)$ at $x=7\%$, with the gap function \cite{prozorov_magnetic_2006}
\begin{equation}\label{gap}
\Delta(T)=\Delta(0)\tanh\left(\frac{\pi T_c}{\Delta(0)}\sqrt{a\left(\frac{T_c}{T}-1\right)}\right) 
\end{equation}
where $a$ is a free parameter that characterizes the rise of $\rho_s(T)$ below $T_c$. This full single gap fit gives $\Delta(0)=2.0k_BT_c$ and $a=1.7$, and, as shown in figure 1a, adequately describes the measured temperature dependence for $x = 7\%$. 

Due to the steeper $\Delta\lambda_{ab}(T)$ away from OptD, the full single gap fit does not work at other dopings. The low temperature variation can be described by a power law, but the full temperature dependence is also well described by a two-gap fit with one gap being small. In cases where the SSS and MFM data do not agree or where only one is available, only one of the fits is possible. We use a phenomenological two-gap fit, $\rho_s(T)=p\rho_1(T)+(1-p)\rho_2(T)$ (Eq.~2), where $\rho_1$, $\rho_2$ are given by Eq.~1 with gaps $\Delta_1$, $\Delta_2$. We fix $T_c$ of the two gaps to be identical and set $a$ for the smaller gap to be 1 to minimize the number of free parameters.  Best fit values are shown in Table~\ref{table_para}. In the power law description,  $\Delta\lambda_{ab}(T)=AT^n$ (Eq.~3).  The best fit value of $n$ varied from 2.3 to 2.9, but the error bars for all data sets are consistent with $n=2.5$.  We therefore fixed $n=2.5$ for simplicity. As shown in Table~\ref{table_para}, the fitted magnitude of $A$ increases away from OptD.  
\begin{table} 
\caption{\label{table_para} Fit parameters for curves in Fig.~\ref{fig_penetration} and \ref{fig_uniformity}, for the two-gap (Eq. 2) and power-law (Eq. 3) models. Power law fits are based on SSS data and two-gap fits on MFM or combined MFM and SSS data. Where both power-law and two-gap fits are shown, SSS and MFM data on different samples agreed. At $x=4.9\%$, 5.1\% and 7\%,  $\Delta\lambda_{ab}(T)$ at multiple positions separated by at least 100 microns are identical. For x=11\%, SSS results from 3
positions are shown. $T_c$ measured on the same sample by MFM or SSS are also provided.} 
\begin{tabular}{l c c c c c c c} \hline \hline
 &  & \multicolumn{4}{c}{full gap}  & & $A$ \\ \cline{3-6} 
$x$ & $T_c(K)$ & $\frac{\Delta_1}{k_BT_c}$ & $\frac{\Delta_2}{k_BT_c}$ & $a$ & $p$ &  & (nm/K$^{2.5}$)\\
\hline $4.9\%$-1& 15.8 & 2.2 & 0.6 & 1.6 & 0.90 & & -\\
$4.9\%$-2  & 15.5 & - & - & - & - &  & 0.26 {\scriptsize (3 positions)}\\
$5.1\%$  & 18.6 & 2.3 & 0.7 & 1.6 & 0.90 &  & 0.09 {\scriptsize (4 positions)}\\
$7\%$ & 22.4 & 3.3 & 1.3 & 1.7 & 0.70 &  & 0.02 {\scriptsize (4 positions)} \\
$8.5\%$ & 19.6& 1.9 & 0.6 & 1 & 0.92 & & -\\
$11\%$  & 13.5 & 1.7 & 0.6 & 0.9 & 0.87 & & 0.38 {\scriptsize (position 1)}\\
$11\%$ & - & - & - & - & - &  & 0.27 {\scriptsize (position 2)} \\
$11\%$  &- & - & - & - & - &  & 0.18 {\scriptsize (position 3)}\\
\hline \hline
\end{tabular}
\end{table}

The third trend we observe is that near $T_c$,  $\rho_s(T)$ of the OptD and UD rises faster with decreasing temperature than weak-coupling BCS theory, MgB$_2$ or OD (Fig.~\ref{fig_penetration}b). This feature is characterized in the full-gap fits: $a$ at $x=5.1\%$ and $7\%$ are significantly larger than the BCS value $a=1$, suggesting a steeper rise of $\rho_s(T)$ with cooling than the weakly-coupled BCS would give. The same feature is shown in $\rho_s(T)/\rho_s(0)$ vs $T/T_c$ plots when comparing with other superconductors. The curves at $x=5.1\%$ (UD) and 7\% (OptD) have similar slopes near $T_c$. Both are steeper than that of MgB$_2$, a weakly coupled two-gap BCS superconductor; or $x=11\%$ (OD).

Local probes allow us to measure at multiple positions on multiple samples and to examine real-space inhomogeneity. On all samples we measured with MFM, we obtain identical $\lambda_{ab}(T)$ and $\Delta\lambda_{ab}(T)$ within errors on positions several microns apart.  On all samples studied with SSS, we measured three or four positions separated by hundreds of microns. For the $x=4.9\%$, $5.1\%$, and $7\%$ samples, the $\Delta\lambda_{ab}(T)$ measurements were identical within errors at each location. We observe spatial variation in $\Delta\lambda_{ab}(T)$ on one $x=11\%$ sample as shown in Fig.~\ref{fig_uniformity}, resulting in the variations in fit parameters shown in Table~\ref{table_para}. In principle, such spatial variations could come from doping inhomogeneity in the sample, or from roughness of the sample surface, or from other variations. In order to quantify doping inhomogeneity in the sample, we performed x-ray microanalysis after the SSS measurement was complete.  The 3 $\mu$m beam was scanned across a 1.2 mm long line in 20 $\mu$m steps. We observe no systematic change in the doping level to within $\pm2\%$ of the total Co concentration. We do not think the spatial variation comes from sample surface roughness because we checked the topography by susceptometry scans in-situ and by optical microscopy after the measurement, and because roughness is expected to overestimate $\Delta\lambda$, while the three values are either smaller than or the same as the MFM result.

The systematic trends that we have observed in $\rho_s(T)$ should be considered in light of properties in the superconducting states that are expected to evolve with doping, including the structure of the gap in $k$-space on multiple bands \cite{Muschler_Band_2009, Liu_NatPhys_2010}, magnetic scattering, other forms of scattering \cite{Gordon_Pene_pairbreaking_2010}, and transfer of spectral weight to spin fluctuations and the magnetic phase \cite{Inosov_spindynamics_2010}. In particular, the three trends of $\rho_s(T)$ can be accounted for by the interplay between magnetism and superconductivity.

The first trend, $\rho_s(0)$ dropping more rapidly on moving towards UD than towards OD, follows naturally from the fact that the structural and magnetic transitions lead to significant Fermi surface reconstruction \cite{Analytis_Quantum_2009, Liu_NatPhys_2010}, resulting in smaller electron and hole pockets and therefore fewer charge carriers for the superconducting state. In the UD cuprates, the reduction of $\rho_s(0)$ approximately following $T_c$ \cite{Uemura_universal_1989} has been often attributed to phase fluctuations of the superconducting state \cite{emery_importance_1995}. We observed an even faster drop of $\rho_s(0)$ than $T_c$, consistent with the scenario that coexisting order, e.g. magnetic order, removes a large number of itinerant carriers that might otherwise enter the superconducting condensate.

The second trend, weakening of fully-gapped behavior away from OptD, agrees with heat transport measurements which have also reported an increase in low-energy quasi-particle excitation on either side of OptD \cite{Reid_caxisheat_2010}. The observation indicates strong pair-breaking scattering or anisotropic superconducting gap structure in the $s\pm$ pairing symmetry \cite{vorontsov_superfluid_2009}. The Fermi surface reconstruction resulting from the magnetic order on the UD side is not expected to lead to nodes, but may result in deep minima in the gap structure \cite{Parker_Coexistence_2009, Ghaemi_Anomalous_2010}. The increasing strength of the static order and low-frequency magnetic fluctuations on the UD side \cite{Lester_Neutron_2009, Pratt_Coexistence_2009} could enhance pair-breaking magnetic scattering \cite{Monthoux_spinfluctuation_1994}, giving rise to a power law dependence in $\Delta\lambda_{ab}(T)$ \cite{vorontsov_superfluid_2009} that gets sharper with less doping on the UD side. On the OD side, deep gap minima may result from an anisotropic reduction of pairing strength as the doping moves further from the static magnetic order. Although pair-breaking may play some role and is one possible explanation for the spatial variation in the 11\% sample, two facts suggest that our results are not dominated by pair-breaking processes from sample imperfection: the doping dependence we report is consistent with the low energy excitations measurements on annealed crystals \cite{Gofryk_effect_2010}, and we observe flat $\Delta\lambda_{ab}(T)$ at OptD.

The third trend, the rapid increase $\rho_s(T)$ of the UD and OptD when cooling through $T_c$, also agrees with the importance of magnetism. If the pairing is mediated by spin-fluctuations \cite{mazin_unconventional_2008}, forming superconductivity pushes the fluctuation spectrum to higher frequency, which further strengthens pairing, leading to a more rapid rise of $\rho_s(T)$ than the standard BCS expression would give \cite{Monthoux_spinfluctuation_1994}. On the OD side, away from the magnetic order, the absence of low-frequency magnetic fluctuations may contribute to the slow rise of $\rho_s(T)$ when cooling through $T_c$. The slower rise at $x=11\%$ than the weak-coupling BCS result may hint that the two gaps have different $T_c$'s. 

To conclude, by locally measuring $\lambda_{ab}(T)$ and $\Delta\lambda_{ab}(T)$, we observe systematic doping evolution of both the zero temperature value and the temperature dependence of $\rho_s(T)$ in  Ba(Fe$_{1-x}$Co$_x$)$_2$As$_2$ single crystals. Using local scanning techniques, we reduce the error from sample inhomogeneity. The three systematic trends we observe on $\rho_s(T)$ across the superconducting dome are consistent with the role of magnetism as a coexisting and competing order to the superconductivity as well as the pairing glue. 

Acknowledgment: This work is supported by the Department of Energy, Office of Basic Energy and Sciences under contract DE-AC02-76SF00515. We thank S.~A.~Kivelson, D.~J.~Scalapino, and M.~R.~Beasley for helpful discussions.

\newcommand{\noopsort}[1]{} \newcommand{\printfirst}[2]{#1}
  \newcommand{\singleletter}[1]{#1} \newcommand{\switchargs}[2]{#2#1}

\end{document}